\input{aipcheck}

\documentclass[
    ,final            
  ]
  {aipproc}

\layoutstyle{6x9}

\newfont{\header}{rptmbi scaled 1600}

\begin{document}

\title{Some Recent Peculiarities of the Early Afterglow}

\author{Tsvi Piran}{
address={Racah Institute for Physics, The Hebrew University,
Jerusalem, 91904, Israel}}

\author{Ehud Nakar}{
address={Racah Institute for Physics, The Hebrew University,
Jerusalem, 91904, Israel}}

\author{Jonathan Granot}{
address={Institute for Advanced Study, Einstein Drive, Princeton,
NJ 08540, USA}}

\begin{abstract}
We consider some recent developments in GRB/afterglow
observations: (i) the appearance of a very hard  prompt component
in GRB 941017, and (ii) variability in the early  afterglow light
curves of GRB 021004 and GRB 030329. We show that these
observations fit nicely within the internal-external shocks model.
The observed variability indicates that the activity of the inner
engine is more complicated than was thought earlier and that it
involves patchy shells and refreshed shocks. We refute the claims
of Berger et al. and of Sheth et al. that the radio and mm 
observations of GRB 030329 are inconsistent with refreshed shocks.
\end{abstract}

\maketitle


\subsection{Introduction}

 The early afterglow and the
GRB/afterglow transition are among the unexplored regimes of GRBs.
A great progress on this front was achieved during the last year
when, following  fast HETE II identifications, two afterglows
(GRBs 021004, 030329)
were followed from very early on showing remarkable variability
and rich structure. Also, somewhat unexpectedly, a search within
the BATSE/EGRET archives revealed a new very hard long lasting
($\sim 200\;$s) component of GRB 941017.  This is most likely a
manifestation of the early afterglow and of the GRB/afterglow
transition. We discuss these developments and their implication
within the internal-external shocks model.

\subsection{The Prompt High Energy Emission from GRB 941017 as a
GRB/afterglow transition}

Recently, Gonz\'alez et al. \cite{Gonzalez03} discovered a high
energy tail that extended up to $200\;$MeV in the combined BATSE
and EGRET data of GRB 941017. The tail had a hard
spectral slope ($F_\nu \propto \nu^0$) up to $200\;$MeV. It
appeared $\sim 10$-$20\;$s after the beginning of the burst and
displayed a roughly constant flux while the lower energy component
decayed. At late times ($\sim 150\;$s after the trigger) the very
high energy ($\sim 10$-$200\;$MeV) tail
 had a luminosity $\sim 50$ times higher than the ``main''
$\gamma$-ray energy band ($\sim 30\;$keV-$2\;$MeV).

Granot \& Guetta \cite{GG03} where the first to suggest that we see
here another manifestation of the very early afterglow.
Sari \cite{Sari97} has shown that for long bursts the external
shocks (and hence the afterglow) begin $\sim R/c \Gamma^2$ after
the beginning of the burst while the internal shocks are still
going on and hence the burst is still active. This behavior was
seen in the transition from the harder initial burst to the softer
early afterglow \cite{BureninEtal99,GiblinEtal99,Piro00}. It was
also seen in the prompt optical flash of GRB 990123 where the
lower (optical) energy component did not trace the $\sim$MeV
$\gamma$-rays and a pronounced hard to soft evolution was seen in
the $\gamma$-ray signal.

The very high energy of this emission suggests that it  is inverse
Compton. There are two relevant emitting regions in the early afterglow:
the forward shock and the reverse shock. With typical parameters
\cite{SariPiran99b} the
energy of the synchrotron photons and the electrons' Lorentz
factor in the forward and reverse shocks are:

$${\begin{array}{ccc}
\nu_{\rm synch,F} \approx 0.1\;{\rm MeV}\;(\Gamma/300)^4 ~~; &
\gamma_{e,{\rm F}} \approx 10^4 (\Gamma/300) \\
  \nu_{\rm synch,R} \approx 1\;{\rm eV}\;(\Gamma/300)^2 \qquad ;&
  \gamma_{e,{\rm R}} \approx 300 \\
\end{array}}
$$

There are four possible combinations of seed photons and
scattering electrons:
$${\begin{array}{ccc}
   &{\rm Rev. ~ shock ~electrons }& {\rm For. ~ shock ~electrons}
   \\ \nonumber
  {\rm  Rev.~ shock ~ photons }& \sim 0.1\,{\rm MeV} (\Gamma/300)^2 &
  \sim 100\,{\rm MeV} (\Gamma/300)^3\\
  {\rm For. ~ shock ~ photons} & \sim 10\,{\rm GeV}(\Gamma/300)^4 &
  \sim 10\,{\rm TeV} (\Gamma/300)^5 \;{\rm (within ~Klein~ Nishina)}\quad \\
\end{array}}
$$
While these approximate results are very sensitive to $\Gamma$,
they indicate  that inverse Compton scattering of the reverse
shock photons on the forward shock electrons yeilds the right
energy range. The detailed calculations of Pe'er \& Waxman
\cite{PeerWaxman03}   confirm these naive estimates. The main
problem, however, is not to explain the location of the spectral
peak but to explain the spectral slope ($F_\nu \propto \nu^0$) and
the temporal slope ($F_\nu \propto t^0$). Pe'er \& Waxman
\cite{PeerWaxman03} reproduce the spectral slope
by requiring that the synchrotron self absorption frequency of the
reverse shock emission would be high enough to effect the observed
spectrum. 
Granot \& Guetta \cite{GG03} reproduce both the temporal and
spectral behaviors by considering a slightly different scenario
where the high energy component is produced by Synchrtron self
Compton within the reverse shock. They require a slightly higher
external density 
with a somewhat unusual profile, $\propto R^{-1}$,  for a uniform
ejecta shell (which may explain the rareness of the event).
In both models the high energy component is a clear manifestation
of the onset of the afterglow and the GRB/afterglow transition.

\subsection{Afterglow Light curve variability}

{\header Theory:} The different scenarios that lead to afterglow
temporal variability 
can be distinguished according to their characteristic features.
Density variations produce only weak fluctuations above the
cooling frequency, $\nu_c$, and cannot produce sharp changes in
the light curve. Energy variations produce variability both above
and below $\nu_c$, and can arise either due to refreshed shocks or due to a
patchy shell structure. These two mechanisms produce very
different light curves. While the former produce a step-wise
increase in the light curve, the later produces random
fluctuations with a decreasing amplitude.

Rees \& M\'esz\'aros \cite{ReesMeszaros}, Kumar \& Piran
\cite{KumarPiran00a} and Sari \& M\'esz\'aros \cite{SariMeszaros}
suggested that slow shells take over the slowing down matter
behind the afterglow shock and produce {\it refreshed shocks}.
Slow shells with $\Gamma_s$ emitted from the source right after
the fast ejecta, catch up and collide with the slowing down ejecta
at an observer time $t \sim 0.25
(\Gamma_s/10)^{-8/3} (E_{\rm iso,52}/n_0)^{1/3}\;$days (where
$E_{\rm iso,52}$ is the isotropic kinetic energy in units of
$10^{52}\;$ergs and $n_0$ is the external density in cm$^{-3}$)
when the ejecta's Lorentz factor drops slightly below $\Gamma_s$.
The clearest feature of refreshed shocks is a monotonous increase
in the overall energy. Therefore the observed flux  can only
increase (relative to the expected decay). The light curve has a
step wise form with each step produced by the arrival of a single
shell. This step wise structure is seen both above and below the
cooling frequency with a similar amplitude. Each step (in the
optical light curve) should be accompanied by a flare in low
frequencies
that is produced by the reverse shock which propagates back into
the slow shell \cite{KumarPiran00a,SariMeszaros}. The time scale,
$\Delta t$ of the steps and the corresponding flares depends on
their timing relative to the jet break. Before the jet break the
refreshed shocks are ``locally'' spherically symmetric and
therefore the angular time imposes $\Delta t \sim t$
\cite{KumarPiran00a}. The intensity of the reverse shock flare in
this regime is calculated in \cite{KumarPiran00a}  , and the decay
after the peak is $\propto t^{-2}$ \cite{SariPiran99b}.  In the
post-break case the cold slow shells may not expand sideways (if
cold enough). Then they  keep their original  angular size,
$\theta_j$, which is smaller than $\Gamma_s^{-1}$. In this case
$\Delta t \approx t_j < t$, where $t_j$ is the jet break time
\cite{GNP03} and the transition is fast. A reverse shock flare is
expected in this case as well. However, its frequency, intensity
and temporal decay (which is expected to be steeper than in the
spherical case) are much harder to calculate.

Kumar \& Piran \cite{KumarPiran00b} suggested, in the {\it Patchy
shell model}, that the shells  have an intrinsic angular
structure. As the blast wave decelerates the angular size of the
observed region ($\sim 1/\Gamma$) increases. The effective
(average) energy of the observed region and hence the observed
flux, relative to the expected decay, varies with time depending
on the angular structure. The variability time scale is $\Delta t
\sim t$ \cite{NakarOren}. The averaging over a larger and larger
random structure leads to a decay of the envelope as $t^{-3/8}$
\cite{NakarOren,NPG03}. An important feature of this scenario is
the break of the axial-symmetry and therefore the production of a
linear polarization. The variation of the polarization, both in
degree and in angle, are correlated with the light curve
variations \cite{NakarOren,GK03}. The variability will be observed
both above and below the cooling frequency $\nu_c$ with a similar
amplitude.

Wang \& Loeb \cite{WangLoeb},  Lazzati et al. \cite{Lazzati02} and
Nakar et al. \cite{NPG03} considered {\it External density
variations}. Such  variations may result from ISM turbulence or
from a variable pre-burst stellar wind. Wang \& Loeb
\cite{WangLoeb} analyzed the light curve resulting from mild
density fluctuations due to ISM turbulence. They show that these
density fluctuations can produce short time scale ($\Delta t <
0.1t$) and low amplitude ($\sim 10\%$) fluctuations in the light
curve. Lazzati et al.\cite{Lazzati02}, Nakar et al. \cite{NPG03}
and Nakar \& Piran \cite{NP03} considered large amplitude
spherical density fluctuations. A basic feature of the resulting
light curve that distinguishes it from energy variations is that
in the former the light curve is different above and below the
cooling frequency, $\nu_c$. Density variations produce only weak
fluctuations above $\nu_c$. The amplitude of the fluctuations
above $\nu_c$ is at most tens of percents and it is much smaller
than the amplitude of the fluctuations below $\nu_c$ \cite{NP03}.

A second feature of density fluctuations is their inability to
produce a sharp variation (either increase or decrease) in the
light curve \cite{NP03a}. First, we note that because of angular
spreading, spherical density drops cannot produce decays sharper
than $t^{-2.6}$ \cite{NP03,KumarPanaitescu} and even this decay is
reached very slowly. More interesting is the fact that even a
sharp density enhancement cannot produce a steep increase in the
light curve. The earlier calculations \cite{NPG03,Lazzati02,NP03}
assumes that the ejecta can be described by a Blandford-McKee
solution whose density profile varies instantaneously according to
the external density. These calculations do not account, however,
for the reverse shock resulting from density enhancement and its
effect on the blast-wave. Thus the above models are limited to
slowly varying and low contrast density profiles. Now, the
observed flux depends on the external density, $n$, roughly as
$n^{1/2}$. Thus, a large contrast is needed to produce a
significant re-brightening. Such a large contrast will, however,
produce a strong reverse shock which will sharply decrease the
Lorentz factor of the emitting matter behind the shock,
$\Gamma_{sh}$, causing a sharp drop in the emission below $\nu_c$
and a long delay in the arrival time of the emitted photons (the
observer time  is $\propto \Gamma_{sh}^{-2}$). Both factors
combine to suppresses the flux and to set a strong limit on the
steepness of the re-brightening events caused by density
variations. Note that while non spherical density fluctuations may
lead to a steeper decline they  usually do not
lead to a steeper increase in the flux.

{\header Implications:} The early afterglow of {\bf GRB 021004}
showed clear deviations from a smooth power law decay, lasting
from 0.04 days to 3 days. The fluctuations in the light curve
where accompanied by fluctuations  both in the degree and in the
angle of the polarization (\cite{NakarOren} and references
therein).

The steep decays after each bump  imply that the variations do not
result from refreshed shocks. Thus,  variable external density
variations \cite{NPG03,Lazzati02,HeylPerna} and the patchy shell
model \cite{NakarOren,NPG03} were considered as possible
explanations. Unfortunately, the X-ray observations are not
detailed enough to clearly distinguish between density and energy
variations
(although the former are favored by \cite{HeylPerna}). However,
the sharp decays cannot be produced by ``locally'' spherical
density variations \cite{NP03}. Furthermore, the first bump
requires, using the instantaneous Blandford-McKee approximation,
an increase in the external density 
by a factor of $\sim 10$ over $\Delta R/R \approx 0.05$
\cite{Lazzati02,NP03}.
Such a density contrast produces a mildly relativistic reverse
shock which reduce $\Gamma_{sh}$ by a factor of $\approx 2$,
making the approximation inconsistent. Preliminary results
\cite{NP03a} of detailed numerical simulations (including both
hydrodynamics and synchrotron radiation) suggest that this bump
cannot be produced by density enhancement (due to the
suppressed  forward shock emission, caused by the reverse shock).
These results leave the patchy shell as the only viable
explanation. Indeed, Nakar \& Oren \cite{NakarOren} show that
patchy shell can reproduce the light curve (including the sharp
rise and steep decay of the first bump). They show that angular
energy profiles which produce the observed light curve produce
also a polarization curve that fits the observed polarization. We
conclude that angular energy fluctuations are the dominant process
that produce the observed fluctuations in GRB 021004.

In addition to the  remarkable supernova signature, the optical
afterglow {\bf GRB 030329} has shown also a unique variability
(\cite{GNP03} and references therein). Several step-wise bumps, at
$t=1.5, 2.6, 3.3$ and $5.3\;$days, are seen after the jet break
at $t_j\approx 0.5\;$days. The first bump was the largest and best
monitored, but even with the less dense monitoring of the later
bumps the step-wise profile (where after each bump the original
decay slope is resumed) is clear. All the bumps had a short rise
time $\Delta t \approx 0.4$-$0.8\;{\rm d} < t$. The step-wise
profile seems like a clear signature of post-break refreshed
shocks \cite{GNP03}, where $\Delta t \approx t_j < t$ because the
later slower shells did not expand sideways before colliding with
the faster earlier ejecta. Moreover, the energy injected in these
shocks is 10 times the energy in the original blast-wave. This
late (or rather slow) energy injection explains an additional
peculiarity of this GRB: the low energy output in $\gamma$-rays
and in the early X-ray afterglow.

Berger et al. \cite{Berger03} suggested that the first bump and
the energy deficiency can be explained by a two component jet: A
slow and energetic component with a wide half-opening angle
($17^\circ$) dominates the afterglow {after $1.5\;$days}, and a
fast component with a narrow half-opening angle ($5^\circ$)
dominates the
afterglow before $1.5\;$days. The slow component is observed only
after $1.5\;$days$ \approx t_{\rm dec}\approx 0.5
(\Gamma_s/10)^{-8/3} (E_{\rm iso,52}/n_0)^{1/3}\;$days, since only
at this time its reverse shock consumes the slow shell. However,
this model, which received a great publicity, predicts that the
rise time, $\Delta t$, of the first bump should be of the same
order as the observed time, $t_{\rm dec}$. Furthermore, it
predicts a smooth light curve after the first bump. Both
predictions are contradicted by the observations.

Millimeter observations of GRB 030329 \cite{Sheth} show that at
100 \& 250 GHz the flux is rather constant during the first week.
Most surprising are two measurements at $100\;$GHz, one before the
first bump ($0.6$-$1\;$d) and one after ($1.7$-$1.9\;$d), which
show a constant flux. Sheth et al. \cite{Sheth} and Berger et al.
\cite{Berger03}  claim that these results support the two
component jet model and reject the refreshed shocks model due to
the lack of radio flares. However this analysis overlooks the fact
that the encounter of the slow component in the two-components jet
model with the external matter produces 
a reverse shock which should produce a radio flash. This flash is
analogous to the optical flash produced by the deceleration of the
fast component \cite{SariPiran99b}, and to the radio flare
expected in the refreshed shocks model. The timing of this flash
is $t_{\rm dec}=1.5\;$d and its magnitude is easily calculated.
The contribution from this reverse shock at $t_{\rm dec}=1.5\;$d
is larger by a factor of  up to $\Gamma_s^{5/3} \sim 15$ than the
flux from the forward shock. Thus a very bright and fast fading
millimeter flash is expected in the two component jet model.  A
more detailed calculation (using \cite{SariMeszaros} and the
parameters of the wide jet presented in \cite{Berger03}) shows
that both $\nu_a$ and $\nu_m$ of the reverse shock at $t_{\rm
dec}$ are around 100-200$\;$GHz and that the flux at $\nu_m$ is
$\sim 500\;$mJy which is an order of magnitude larger  than the
expected flux from the forward shock and than the observed fluxes
at 100 \& $250\;$GHz ($\sim 50\;$mJy) at this time.

The main argument of \cite{Berger03} in favor of a two-components
jet is the existence of a second jet break in the radio after
$t_{j,2}\sim 10\;$d. However, even this argument is not strongly
supported by the data. According to this model the flux below
$\nu_m$ should rise as $t^{1/2}$ before $t_{j,2}$
and decay as $t^{-1/3}$
after $t_{j,2}$. However, the data of \cite{Sheth} contradict this
prediction. At 100 \& $250\;$GHz the flux is constant before the
passage of $\nu_m$ and the decay after this passage (at $t=6$ \&
$8\;{\rm d}<t_{j,2}$, respectively)
is steeper than $t^{-1.7}$ in both bands. This looks like a clear
signature of a post break behavior in the radio at $t>1\;$d. Now
the radio observations at lower frequencies ($\leq 22\;$GHz) do
not conform with the simple post-break model ($F_{\nu}\propto
t^{-1/3}$). The flux at these frequencies rises with time before
the passage of $\nu_m$. Interestingly enough, this radio behavior
is exactly the one predicted as the post break radio behavior by
\cite{Granot01}, using a 2D relativistic hydrodynamical
simulations. We find the striking similarity between the radio
observations and Fig. 2 of \cite{Granot01} as a very strong
support that the broad band
data are totally consistent with a single jet.

Sheth et al., \cite{Sheth} emphasize the fact that radio flare
\cite{KumarPiran00a} was not detected during the first bump at $t
\sim 1.5$ days and argue that this rules out the refreshed shocks
model. However, in  the post-break refreshed shocks model, the
radio flash is expected to be fainter and to decay faster than the
radio flash in the two-components jet model (due to the lower
energy and the lateral spreading of the slow shell as opposed to
the more energetic and ``locally'' spherical wide jet). Thus, by
assuming spherical symmetry,  Sheth et al., \cite{Sheth} over
estimate the expected flash in the post-break refreshed shocks
scenario. It may be that the flash was missed by the sparse
measurements due to its lower intensity and faster decay. A
detailed (and highly non trivial) calculations should be done in
order find out.

\subsection{Summary}

With an increasing flow of new observations we discover that GRBs
and afterglows are richer  than what was previously thought. The
simple spherical theory had to be modified, first with jets and
now with additional angular structure (patchy shells) and more
extended velocity structure (refreshed shocks). The simple
synchrotron theory has to be modified with inverse Compton
scattering. One can worry, are we adding epicycles trying to
revive a wrong theory? We don't believe so. Complications and
variation are common in astrophysics and are found everywhere in
nature. Moreover, the three main themes that have been introduced
here: patchy shells, refreshed shocks and inverse Compton, were
not invoked aposteriori to explain the new observations. On the
contrary, all three have been suggested long ago. It is just
natural and  even reassuring to discover them when better data
become available.

\vspace{0.25cm}

The research was supported by US-Israel BSF.

\end{document}